...............................................................................

\magnification=1200
\baselineskip=20 pt

\line{\hfil MRI-PHY-96-21}
\line{\hfil hep-ph/9607227}

\def\mt{m_t}
\def\rb{R_b}
\def\delrb{{\delta R_b}}
\def\delgbl{{\delta g^b_l}}
\def\delgbr{{\delta g^b_r}}
\def\gblsm{(g^b_l)_{sm}}
\def\gbrsm{(g^b_r)_{sm}}
\def\delgtl{{\delta g^t_l}}

\def\gel{g_{E,L}}
\def\guer{g^U_{E,R}}
\def\gder{g^D_{E,R}}
\def\gmmuu{{\gamma^{\mu}}}
\def\gmmud{{\gamma_{\mu}}}
\def\delftv{{\delta F^t_v}}
\def\delfta{{\delta F^t_a}}
\bigskip
\centerline{\bf Implications of $m_t$ and $\rb$  on $Zt\bar {t}$}

\centerline{\bf couplings in standard ETC models}
\vskip .4truein
\centerline{\bf Uma Mahanta}
\centerline{\bf Mehta Research Institute }
\centerline{\bf 10 Kasturba Gandhi Marg }
\centerline{\bf Allahabad-211002, India}
\vskip 1truein
\centerline{\bf Abstract}

In standard ETC models the sideways and diagonal ETC interactions
contribute to $\delrb $ with opposite signs. The aim of this article
is to study the implications of the CDF value for $m_t$ and the LEP
value for $R_b$ on $zt \bar {t}$ couplings where the LH sideways and
diagonal ETC effects interfere constructively. We find that for 
$m_t=175 $ Gev, $\delrb =.0022 $ and $m^2_s=m^2_d$,  $F^t_v$ and
$F^t_a$ are modified by 19\% and 7\% respectively from their SM
values. The constrains implied by these deviations on diagonal ETC
 scenarios and the feasibility of probing them at NLC
  through polarization and angular distribution studies in
$e^+e^-\rightarrow t\bar {t}$ are also considered.
              
\vfill\eject

In standard ETC models the large mass ($m_t\approx 175$ Gev) of the top
quark is presumably due to sideways ETC dynamics (that connect ordinary
fermions to technifermions) at relatively low energy scales
($\approx 1$ Tev) [1].
Because of $SU(2)_L$ gauge invariance of ETC interactions the same sideways
ETC dynamics also gives rise to a sizeable neagtive shift  ($1.8\%$) in $R_b$
[2] that can be detected with the present LEP precision [3]
in measuring $R_b$
($R_b^{exp}\approx .2178\pm .0011 $). On the other hand diagonal ETC
interactions (between a pair of technifermions or a pair of ordinary
fermions)
 give rise to a positive correction to $R_b$ [4]. The overall
contribution to $\delrb $ can therefore be of either sign and it can be
large or small depending upon the relative size of the sideways and
diagonal contributions. In contrast the sideways and diagonal ETC
interactions interfere constructively in $\delgtl$.
Hence it is possible for a low enough ETC scale, that both contributions to
$\delgbl$ are individually quite large in magnitude but their difference is
small so as to fit the observed $\delrb$ which is at the level of a few
percent only. Such a scenario would produce large deviations from the SM
 in the LH $zt\bar {t}$ couplings.
The aim of this article is threefold:
i) to investigate if the recent experimental values
of $m_t$ and $R_b$ imply large corrections to $g^t_l$ and $g^t_r$ 
 in standard ETC models ii) the constraints  imposed by 
these deviations on
the unknown parameters of the model and iii) the feasibility of probing the
deviations at NLC.

To illustrate our point we shall consider the one family TC model of 
Appelqusit and Terning [5].
 For simplicity the TF's will be assumed
to be in the fundamental representation of an $SU(N)_{TC}$ gauge group.
It can be shown that in this model the sideways ETC gauge boson exchange
gives rise to the following four-fermion Lagrangian [4]

$$\eqalignno{L^s_{4f}&=-{( \gel )^2\over 2m^2_s} \bar{Q}_L\gmmuu\psi_L
\bar{\psi}_L\gmmud Q_L -{( \guer )^2\over 2m^2_s} \bar{U}_R\gmmuu t_R
\bar{t}_R\gmmud U_R\cr
&-{ ( \gder )^2\over 2m^2_s }\bar{D}_R\gmmuu b_R \bar{b}_R\gmmud D_R.
&(1)\cr}$$  

On the other hand the diagonal ETC gauge boson gives rise to the four
fermion Lagrangian

$$\eqalignno{L^d_{4f}&={1\over 4m^2_d (N_{TC}+1)} (\guer - \gder)
\bar {Q}_R \gmmuu\tau_3 Q_R (\gel\bar {\psi}_L\gmmud\psi_L +\guer
\bar {t}_R\gmmud t_R\cr
&+\gder \bar {b}_R\gmmud b_R).&(2)\cr}$$

Here $\gel$ is the effective ETC gauge coupling to LH fermions.
 $\guer$ ($\gder$) is the effective ETC gauge 
coupling to RH fermions with
$I_3=1/2 $ ($I_3=-1/2 $). We shall assume that the techniquark sector
is intrinsically isospin symmetric i.e. $\langle \bar {U}U\rangle =
\langle \bar {D}D\rangle $.
To obtain the large mass splitting between t and b under this condition
requres that $\guer\gg \gder $. Since spontaneous CSB in the TC sector
occurs only in the I=1 channel, from $L^d_{4f}$ we have dropped those terms 
which contain isospin singlet TF current. Fierz transforming the above
expression for $L^s_{4f}$ both with respect to Dirac and gauge group
indices and dropping terms which contain isospin singlet TF current
we get

$$\eqalignno{L^s_{4f}&=-{ (\gel )^2\over 4m^2_s N_c}\bar {Q}_L\gmmuu
\tau_a Q_L \bar {\psi}_L\gmmud \tau_a \psi_L -{ (\guer)^2\over 4m^2_s 
N_c}\bar {Q}_R\gmmuu \tau_3 Q_R\bar {t}_R\gmmud t_R\cr
&+{ (\gder)^2\over
4m^2_s N_c} \bar {Q}_R\gmmuu\tau_3 Q_R \bar {b}_R\gmmud b_R.&(3)\cr}$$

Below the TC chiral symmetry breaking scale we must replace the TF current by
the appropriate sigma model current [6]. Considering only the
term involving the weak Z boson we get in unitary gauge

$$\bar {Q}_L\gmmuu \tau_3 \otimes 1_3 Q_L=i{ f_Q^2\over 2}Tr(\Sigma^+\tau_3
\otimes 1_3 D^{\mu}\Sigma )_{\Sigma =1}=-{g\over 2c} N_c f_Q^2 Z^{\mu}.
\eqno(4a)$$

$$\bar {Q}_R\gmmuu\tau_3\otimes 1_3 Q_R=i{ f_Q^2\over 2}Tr(\Sigma
\tau_3\otimes 1_3 (D^{\mu}\Sigma)^+)_{\Sigma =1}={g\over 2c}N_c f_Q^2
Z^{\mu}.\eqno(4b)$$

where $1_3$ is the unit operator in color space. The sideways ETC induced
non-standard couplings of t and b to Z boson are therefore given by
$$\eqalignno{L^s_{4f}&={ (\gel )^2\over 8m^2_s} {g\over c} f_Q^2
Z_{\mu}\bar {\psi}_L\gmmuu\tau_3 \psi_L-{(\guer )^2\over 8m^2_s}{g\over c}
f_Q^2 Z_{\mu}\bar {t}_R\gmmuu t_R\cr
&+{ (\gder )^2\over 8m^2_s} {g\over c} f_Q^2 Z_{\mu}\bar {b}_R\gmmuu b_R.
&(5)\cr}$$

The above Lagrangian implies that
$$\delta g^{ts}_L\approx -{(\gel )^2 f^2_Q\over 8m^2_s}
\ \  ,\delta g^{ts}_R \approx {(\guer )^2 f_Q^2\over 8m^2_s}.\eqno(6a)$$

$$\delta g^{bs}_L\approx {(\gel )^2 f_Q^2\over 8m_s^2}\ \ ,\delta
g^{bs}_R\approx -{(\gder )^2 f_Q^2\over 8m^2_s}.\eqno(6b)$$

Similarly for the diagonal ETC exchange we obtain the following
deviations from the SM couplings to Z boson

$$\eqalignno{L^d_{4f}&={ ( \guer-\gder )\over 8m^2_d(N_{TC}+1)}{g\over c}
N_c f^2_Q Z^{\mu}(\gel\bar {\psi}_L\gmmud\psi_L +\guer
\bar {t}_R\gmmud t_R \cr
&+\gder \bar {b}_R\gmmud b_R).&(7)\cr}$$

Hence 
$$\delta g^{td}_L\approx -{ ( \guer-\gder )\over 8m^2_d(N_{TC}+1)}\gel
N_c f^2_Q\ \ ,\delta g^{td}_R\approx -{( \guer-\gder )\over
8m^2_d(N_{TC}+1)} \guer N_c f^2_Q.\eqno(7a)$$

$$\delta g^{bd}_L\approx -{ ( \guer-\gder )\over 8m^2_d(N_{TC}+1)}\gel
N_c f^2_Q\ \ ,\delta g^{bd}_R\approx -{( \guer-\gder )\over
8m^2_d(N_{TC}+1)} \gder N_c f^2_Q.\eqno(7b)$$

Adding the sideways and diagonal contributions separately for the LH and RH
couplings of t we get $\delta g^t_L 
\approx -{(\gel )^2 f^2_Q\over 8m^2_s} -{ ( \guer-\gder )\over 
8m^2_d(N_{TC}+1)}\gel N_c f^2_Q$ and $\delta g^t_{R} \approx 
{(\guer )^2f_Q^2\over 8m^2_s} -{( \guer-\gder )\over
8m^2_d(N_{TC}+1)} \guer N_c f^2_Q $. Similarly for  b we get
$\delta g^b_L 
\approx {(\gel )^2 f^2_Q\over 8m^2_s} -{ ( \guer-\gder )\over 
8m^2_d(N_{TC}+1)}\gel N_c f^2_Q$ and $\delta g^b_{R} \approx 
-{(\gder )^2f_Q^2\over 8m^2_s} -{( \guer-\gder )\over
8m^2_d(N_{TC}+1)} \gder N_c f^2_Q $.  
We shall assume that $N_{TC}=2$ so that the TC contribution to the
S parameter is in agreement with the experimental bounds [4]. From 
$\delta g^t_L$ and $\delta g^t_R$ we can compute the non-standard
vector and axial vector couplings of the top quark to Z boson

$$\eqalignno{\delta g^t_v &\approx -{(\gel )^2f_Q^2\over 16m_s^2}
-{(\guer-\gder )(\gel +\guer )f_Q^2\over 16m_d^2 }\cr
&+{(\guer )^2f_Q^2\over 16m_s^2}.&(9a)\cr}$$
 
$$\eqalignno{\delta g^t_a &\approx {(\gel )^2f_Q^2\over 16m_s^2}
+{(\guer-\gder )(\gel -\guer )f_Q^2\over 16m_d^2}\cr
&+{(\guer )^2f_Q^2\over 16m_s^2}.&(9b)\cr}$$

The non-standard contributions to the
vector and axial vector form factors for $Zt\bar {t}$ 
vertex are therefore given by

$$\eqalignno{\delta F^t_v &\approx {1\over (g^t_v)_{sm}}
[-{(\gel )^2f_Q^2\over 16m_s^2}
-{(\guer-\gder )(\gel +\guer )f_Q^2\over 16m_d^2}\cr
&+{(\guer )^2f_Q^2\over 16m_s^2}].&(10a)\cr}$$

$$\eqalignno {\delta F^t_a &\approx {1\over (g^t_a)_{sm}}
[{(\gel )^2f_Q^2\over 16m_s^2}
+{(\guer-\gder )(\gel -\guer )f_Q^2\over 16m_d^2}\cr
&+{(\guer )^2f_Q^2\over 16m_s^2}].&(10b)\cr }$$

where $(g^t_v)_{sm}={1\over 4}-{2\over 3}s^2$ and $(g^t_a)_{sm}=
-{1\over 4}$. $F^t_v$ and $F^t_a$ are normalized to unity at tree level
in the SM. 
$\delgbl$ and $\delgbr$ affects the precision EW measurements
at the Z pole through $\Gamma_b$ or $R_b$. However, since $\delta \Gamma_b
\propto \gblsm\delgbl + \gbrsm\delgbr$ and $\vert \gbrsm \vert\ll
\vert \gblsm \vert$, we can ignore the effect of $\delgbr$ on $\rb$.
In other words words precision measurements of $R_b$ constrains only 
$\delgbl$ satisfactorily but leaves $\delgbr$ largely unconstrained. 
We shall therefore use the LEP value of $\delta R_b$ to impose constrain
on the ETC contributions to $\delgbl$ only. Since the sideways ETC gauge
boson also contributes to $\mt$, another constrain on the ETC gauge
coupling and and the sideways gauge boson mass will arise from the CDF
$(m_t=176\pm 8\pm 10 Gev)$ or D0 $(m_t=199\pm 20 \pm 22 Gev)$ value 
for $m_t$ [7]. Using naive dimensional analysis [8] and large $N_{TC}$
scaling we can write 
$$m_t\approx -{\gel\guer\langle \bar{U}_LU_R\rangle\over 2m_s^2}\approx
{\gel\guer\over 2m_s^2}4\pi f_Q^3\sqrt{{N_c\over N_{TC}}}.\eqno(11a)$$

The above eqn. implies that $\gel$ and $\guer$ must be of the same sign.
For $m_t\approx 175 Gev$, $\sqrt {3}f_Q\approx 247 Gev$ and $N_{TC}=2$
we get ${\gel\guer f_Q^2\over m_s^2}\approx .1594$. On the other hand
the LEP value of $\delrb$ imposes the following constrain on the
expression for $\delgbl$ in the limit of vanishing $\gder$ (in this limit
$\delgbr$ vanishes)

$${(\gel )^2f_Q^2\over m_s^2}\approx .1594{m_s^2\over m_d^2}-10.3361
\delta R_b.\eqno(11b)$$.

 From (11a) and (11b) we obtain the relation
${\gel\over \guer}\approx {m_s^2\over m_d^2}-64.8438 \delta R_b $. For given 
values of $\delta R_b$ and $m_t$, the ETC contributions
to $\delta F^t_v$ and $\delta F^t_a$ therefore depend only on the 
unknown parameter ${m_s^2\over m_d^2}$. Here we shall consider only
 those values of ${m^2_s\over m^2_d}$ which lie 
  between .5 and 2. Since the LEP value for $R_b$ has been changing
  continuously, we shall treat $\delrb$ as an almost free parameter.
   More precisely we shall calculate $\delftv$ and $\delfta$ for
  $m_t=175$ Gev  and $\delrb =.0011, .0022\ and\  .0044 $. Note that the
difference between the most recent LEP value of $R_b$ and $R_b^{sm}$ 
is .0022.
 We find that for $\delta R_b=
.0011$,\ \  $\delta F^t_v$ $(\delta F^t_a)$ are given by .024(-.084),
-.202(-.077), -.465(-.136) for
 ${m_s^2\over m_d^2}=.5, 1,  and\  2$ 
respectively.
On the other hand if $\delta R_b=.0022$,\ \ 
$\delta F^t_v$ $(\delta F^t_a)$ are given by .056(-.090),
-.194(-.074) and -.460(-.133) 
 for the same set of values of ${m_s^2\over m_d^2}$. Finally for 
 $\delrb = .0044$,\ \  $\delta F^t_v$ $(\delta F^t_a)$ are given by
 .169(-.126), -.179(-.068) and -.448(-.125).

We observe the following features in the ETC contributions to $\delftv$
and $\delfta$.

i)In $\delftv$, the LH sideways  contribution and the diagonal contributions
 (both LH and RH) appear with the same sign (negative). The RH sideways
contribution however appears with opposite sign (positive) relative to the
former. On the contrary in $\delfta$ the sideways contributions (both
LH and RH) and the LH diagonal contribution appear with the same sign
(negative). The RH diagonal contribution however appears with opposite
sign (positive) giving rise to some amount of cancellation.

ii)$\delftv$ is more sensitive to low scale ETC physics than $\delfta$
primarily because $\vert (g^t_v)_{sm}\vert<\vert (g^t_a)_{sm}\vert$.

iii) For a given ${m^2_s \over m^2_d}\ge 1 $, as we increase $\delrb$
both $\delftv$ and $\delfta$  decrease in magnitude. But the change is not
that significant. On the other hand for ${m^2_s \over m^2_d}<1 $,
$\delftv$ increases quite rapidly with increasing $\delrb$. However $\vert
\delfta \vert$ increases only slightly under this condition. The reason
being  for ${m^2_s \over m^2_d}\ge 1 $ the terms that contribute 
constructively  in $\delftv$ dominate and they are not much sensitive
to $\delrb$. On the other hand for ${m^2_s \over m^2_d}< 1 $, the term that
contributes destructively in $\delftv$ dominates and it is quite sensitive
to $\delrb$.

iv)For a fixed $\delta R_b$, as we increase ${m_s^2\over m_d^2}$ both
$\delftv$ and $\delfta$ increase in magnitude. The effect is significant
for both, but it is more dramatic for $\delftv$. This happens because
with decreasing $m_d^2$ the diagonal contribution to $\delta R_b$
increases. To get the same $\delta R_b$, the LH sideways contribution
must therefore increase in magnitude and the two effects interfere
constructively  in $\delftv$ and $\delfta$. 

v)ETC interactions renormalize the $zt\bar {t}$ vertex in such a way that
the strength of axial charge always decreases. On the other hand the
magnitude of the vector charge decreases (increases) if 
${m^2_s \over m^2_d}\ge 1 $ (${m^2_s \over m^2_d}\le .5$) for all
relevant values of $\delrb$. 

Note first that QCD and EW corrections to $F^t_v$ and $F^t_a$ in the context
of the SM are only of the order of a few percent or less. Thus large
corrections ($\ge 10 \% $) to these form factors would imply the presence
of new physics. Second we find that even if $\delrb $ is constrained
to a few percent, the resulting $\delftv $ and $\delfta $ can be greater
than 10\% in standard ETC models particularly if $ m_d^2\le m_s^2 $. 
The main
reason being in $\delrb$ the sideways and diagonal ETC effects
interfere destructively but in $\delftv $ and $\delfta $ they interfere
constructively thereby giving a large effect [9].

The anomalous vector and axial vector couplings of the top quark to the
 Z boson can be probed with high precision at NLC by studying the angular
disribution of different polarization states of $t\bar {t}$ pair.
Barklow and Schmidt [10] performed a tree level study of NLC sensitivity to
these couplings by applying a maximum-likelihood analysis and using all
the information (helicity angles) in $t\bar {t}$ event. The top mass
was set to $m_t=175 $ Gev and the NLC parameters were chosen to be
$\sqrt s=400 $ Gev, an integrated luminosity of 100 fb$^{-1}$ and 90\%
polarization for electrons.
 The full maximum-likelihood
analysis at 95\% confidence level yields an error of 10\% in $F^t_v $
and $F^t_a $. The ETC induced corrections
to   $F^t_v $  discussed in this article
are therefore expected to be within the sensitivity
reach of NLC for most of the natural values of the parameters. In addition
$\delfta$ will also be measurable with the projected NLC sensitivity
 provided $m^2_d<m^2_s$.

From our study we can conclude that once $\delrb$  is measured quite
accurately at LEP, precision measurements of $\delftv$ and $\delfta$
at NLC can be used to put strong constraints on the ratio ${m^2_s\over
m^2_d}$. For example in order that $\delftv$ and $\delfta$ are less than
the projected  NLC precision of .100 for measuring them, 
it is clear that
${m^2_s\over m^2_d}$  and $\delrb$ must be less than 1 and .0022 respectively.
On the other hand both $\delftv$ and $\delfta$ can exceed the 10\%
precision  limit of NLC if ${m^2_s \over m^2_d}\ge 2 $ and
 $0\le\delrb\le .0044$ or  ${m^2_s \over m^2_d}\le .5$ and $\delrb \ge .0044$.
 Note however that ${m^2_s\over m^2_d}$ cannot be much smaller than 1 
 for otherwise ${\gel\over \guer}$ will become too small or
negative. In any case the fact that the SM has been extremely successful
in explaining almost all the collider data so far to a few percent
implies that ${m^2_s \over m^2_d}\ge 1 $ is likely to be excluded by
precision studies of $zt\bar {t}$ couplings at NLC.

 It is important to compare the 
constraints on diagonal ETC scenarios arising from $zt\bar {t}$ vertex
correction with those from $\delta \rho_{new}$. For $\gder=0$ the ETC
induced isospin violating four TF Lagrangian is given by
$L^{ETC}_{\delta \rho}=-{1\over 4N(N+1)}{(\guer )^2\over m^2_d}
\bar {Q}_R\gmmuu T_3 Q_R\bar {Q}_R\gmmud T_3 Q_R$. It then
follows [4] that 
$\delta\rho_{new}={1\over N}{(\guer )^2\over 16m^2_d}f^2_Q$.
For $\delrb =.0011$ and ${m^2_s\over m^2_d}=.5, 1\  and\  2$ 
$\delta \rho_{new} $ is given by .0029, .0027 and .0026. On the other
hand for $\delrb=.0022 (.0044)$  $\delta \rho_{new} $
is given by .0035 (.0058), .0030 (.0035) and .0027 (.0029)
 for the same set of values of $m^2_s\over m^2_d$. 
We find that for a fixed value of $\delrb$ ($m^2_s\over m^2_d$)
$\delta \rho_{new}$ decreases (increases)
 as $m^2_s\over m^2_d$ ( $\delrb$ )
increases. The present experimental bound [11] on $\delta\rho_{new}$ is
$\delta\rho_{new}\le .0040$. This implies that in order to satisfy
the $\delta\rho_{new}$ constraint $\delrb$ must be less than .0022
and $ {m^2_s\over m^2_d}\ge .5$ or ${m^2_s\over m^2_d}$
must be greater than 1 and  $0\le \delrb\le .0044$. 
Comparing the constraints  arising from $zt\bar {t}$ vertex 
correction with those from $\delta\rho_{new}$ we find that
small ($< 10\%$) $zt\bar{t}$ vertex correction and small ($<.0040$)
$\delta\rho_{new}$ can arise simulaneously in  diagonal ETC scenario
 only if $\delrb <.0022$ and ${m^2_s\over m^2_d} <1$. It is clear
 therefore that diagonal ETC scenarios suffer both from large $zt\bar{t}$
 vertex correction and large $\delta\rho_{new}$ problem for most
values of $\delrb$ and ${m^2\over m^2_d}$. It has recently been shown
 [12] that the most dangerous weak-isospin violating effects in realistic
commuting
ETC models arise not from 
diagonal (TC singlet) ETC gauge bosons but from massive
 ETC gauge bosons in the adjoint representation of TC.
  The contribution of these gauge bosons
to $\delta\rho_{new}$ is of order 6\% which exceeds the 
present experimental
bound by more than an order of magnitude. In order to solve the 
$\delta\rho_{new}$ problem in such models, either one has to fine
tune the relevant ETC gauge coupling close to criticality or construct
models that do not contain massive adjoint
 ETC gauge bosons.
 
Acknowledgement: The author would like to thank Dr. Sekhar Chivukula
for careful reading of the preliminary version of the manuscript
and for making useful comments.

\centerline{\bf References}

\item {1.} R. S. Chivukula, K. Lane and A. G. Cohen, Nucl. Phys. B 343,
554 (1990); T. Appelquist, J. Terning and L. C. R. Wijewardhana,
Phys. Rev. D 44, 871 (1991).

\item {2.} R. S. Chivukula, S. B. Selipsky and E. H. Simmons, Phys. Rev.
Lett. 69, 575 (1992).

\item{3.} Talk presented by A. Blondel at the Warsaw ICHEP, July 1996.

\item {4.} N. Kitazawa, Phys. Lett. B 313, 395 (1995); G. H. Wu,
Phys. Rev. Lett. 74, 4137 (1995); K. Hagiwara and N. Kitazawa, Phys. Rev.
D 52, 5374 (19950.

\item {5.} T. Appelquist and J. Terning, Phys. Lett. B 315, 139 (1995).

\item {6.} H. Georgi, Weak Interactions and Modern Particle Theory,
Benjamin-Cummings, Menlo Park, CA (1984).

\item{7.}F. Abe et. al. (CDF collaboration), Phys. Rev. Lett. 74,
2626 (1995); S. Abachi et. al. (DO collaboration), Phys. Rev. Lett.
74, 2632 (1995).

\item {8.} A. Manohar and H. Georgi, Nucl. Phys. B 234, 189 (1984).

\item {9.} The general idea that the sideways and diagonal ETC
contributions
interfere constructively in $zt\bar {t}$ vertex was first pointed out
by Michael Peskin (unpublished).

\item {10.} T. L. Barklow and C. R. Schmidt, in DPF 94: The Albuquerque
Meeting, S. Seidel, ed. (World Scientific, Singapore, 1995); 
G. A. Ladinsky and C. P. Yuan, Phys. Rev. D 49, 4415 (1994).

\item {11.} J. Erler and P. Langacker, Phys. Rev. D 52, 441 (1995).

\item {12.} T. Appelquist, N. Evans and S. B. Selipsky, Phys. Lett. B
374, 145 (1996).

\end